\documentclass[twocolumn,showpacs,floatfix,superscriptaddress,amsmath,amssymb,prl]{revtex4}
\usepackage{mathrsfs}
\usepackage{txfonts}
\usepackage{amssymb}
\usepackage{graphicx}
\usepackage{hyperref}
\usepackage{ulem}
 \usepackage{overpic}
 \usepackage{psfrag}
\newcommand{\be}{\begin{equation}}
\newcommand{\ee}{\end{equation}}

\begin{document}

\title{Realizable spin models and entanglement dynamics
       in superconducting flux qubit systems}

\author{Qian Qian Shi}
\affiliation{Centre for Modern Physics and Department of Physics,
Chongqing University, Chongqing 400044, The People's Republic of
China}

\author{Sam Young Cho}
\email{sycho@cqu.edu.cn}
 \affiliation{Centre for Modern Physics and
Department of Physics, Chongqing University, Chongqing 400044, The
People's Republic of China}

\author{Bo Li}

\affiliation{Centre for Modern Physics and Department of Physics,
Chongqing University, Chongqing 400044, The People's Republic of
China}

\author{Mun Dae Kim}
\affiliation{Korea Institute for Advanced Study, Seoul 130-722,
Korea}

\begin{abstract}
 Realizable spin models are investigated
 in a two superconducting flux qubit system.
 It is shown that a specific adjustment of system parameters
 in the two flux qubit system
 makes it possible to realize an artificial two-spin system
 that cannot be found naturally.
 For the artificial two-spin systems,
 time evolution of a prepared quantum state
 is discussed to quantify quantum entanglement dynamics.
 The concurrence and fidelity as a function of time
 are shown to reveal a characteristic entanglement dynamics
 of the artificial spin systems.
 It is found that the unentangled input state can evolute to
 be a maximally entangled output state periodically
 due to the exchange interactions induced by two-qubit flipping
 tunneling processes while single-qubit flipping tunneling
 processes plays a role of magnetic fields for the artificial spins.
\end{abstract}
\date{\today}
\maketitle

 {\it Introduction.}
 Superconducting qubit systems as one of promising candidates
 have been paid much attentions
 for quantum information processing and computing.
 The tunable superconducting devices have provided
 a variety of possibilities to realize
 quantum spin models that are not findable naturally.
 Recent experiments have shown that
 different types of exchange interactions are observable.
 Particularly, there have been demonstrated
 an Ising type interaction
 in two charge qubits \cite{Pashkin}
 and two flux qubits \cite{Izmalkov,Majer}
 and
 an XY type interaction
 in two phase qubits \cite{Berkley,Steffen}
 as well as superconducting single qubits \cite{ChargeQ,Mooij99,Yu}.
 Moreover,
 such realizations of artificial spin systems
 make possible to observe
 entangled states of two qubits \cite{Pashkin,Izmalkov,Berkley,Steffen}.
 Indeed,
 for the time evolution of states in the experiments of
 charge  \cite{Pashkin} and phase qubits \cite{Berkley},
 a partial entanglement has been observed.
 An experiment of
 a capacitively coupled two phase qubits \cite{Steffen}
 shows
 that higher fidelity for the entanglement
 exhibits in an excited level.
 The higher fidelity is caused by two-qubit tunneling processes
 \cite{KimCho}
 between two qubit states, i.e., flipping both qubits.
 Such a two-qubit tunneling processes contributes
 exchange interactions between the two artificial spins.

\begin{figure}
 \setlength{\abovecaptionskip}{-10pt}
 \begin{center}
 \includegraphics[width=0.4\textwidth,height=0.18\textwidth]{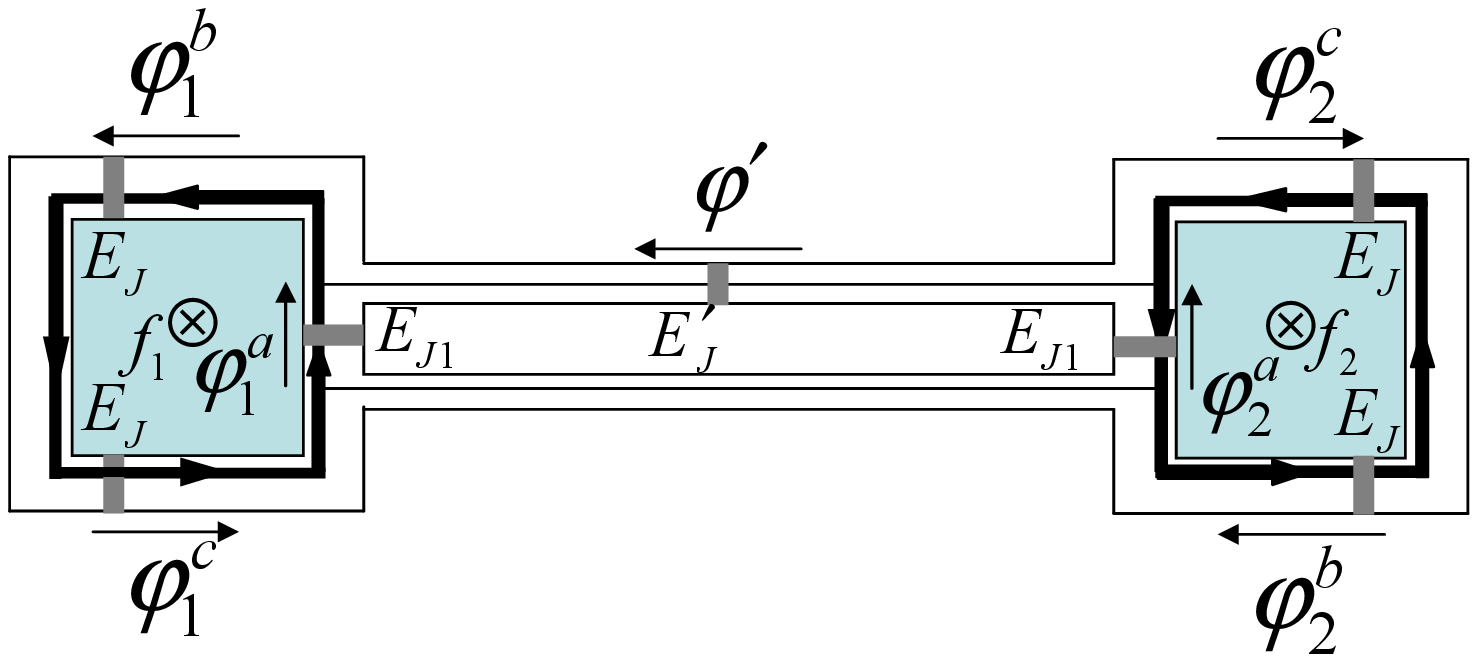}
 \end{center}
\caption{(color online)
 A two flux qubit system. The system is
 composed of two
 (left and right) qubit loops. In order to couple the two flux
 qubits, we use two connecting superconducting wires
 where the Josephson junction $E'_J$
 plays the important role for controlling the interaction between the
 two qubits
 since the two wires give the boundary condition
 as a function of phases $\{\varphi^a_1,\varphi^a_2,\varphi'\}$
 from the fluxoid quantization along the closed path through
 the two connecting wires.
 By varying the amplitude of $E'_J$, the two flux qubit sytem
 can be map into a quantum two-spin model.
 The state of each qubit loop is in a superposed state of
 which $\left|\downarrow\right\rangle$ and
 $\left|\uparrow\right\rangle$ represent the diamagnetic and
 paramagnetic current states, respectively. This schematic of the
 system show the state $\left|\uparrow\uparrow\right\rangle$ that is
 one of possible four states. Here, $\odot$ and $\otimes$ denote the
 direction of the magnetic fields, $f_{1(2)}=\Phi_{1(2)}/\Phi_0$, in
 the qubit loops. $E_{J1}$, $E_{J}$, $E'_{J}$ are the Josephson
 coupling energies of the Josephson junctions in the qubit loops and
 the superconducting connecting wire, and
 $\varphi$'s are phase differences across the
 Josephson junctions.}
  \label{fig:1}
  \end{figure}

 In this paper, we will theoretically investigate
 a possible realization of quantum spin models
 in superconducting flux qubit systems
 by varying a system parameter.
 Especially, we use a phase coupling by introducing
 a connecting wire between the two qubit loops (see Fig.
 \ref{fig:1}) \cite{Kim04}
 because the phase coupling gives
 more controllable parameters than the inductive coupling
 with respect for the manipulation of qubit states \cite{KimCont}.
 It is shown that
 the Josephson junction in the connecting superconducting wires
 plays a role of controller
 in determining exchange interactions between the two qubits.
 In general, it is found that the two flux qubit system can map into
 an XYZ quantum spin model in the presence of magnetic fields.
 We show that specific values of system parameters generate
 various types of quantum spin models.
 Further, to address about time evolution of an input state
 for the two flux qubit system corresponding to quantum spin models,
 we introduce
 the concurrence and fidelity as a function of time
 as a measure of entanglement and evolution of the state.
 It turns out that
 an unentangled (entangled) input state evolves
 to be an entangled (unentangled) state periodically
 with a characteristic period of time.


 {\it Time evolution of quantum states.}
 A system described by two quantum states can be a qubit.
 The two states can be represented in terms of
 pseudo-spin language, i.e., two orthogonal states
 $\left| \uparrow \right\rangle$
 and $\left| \downarrow \right\rangle$.
 Then,
 any normalized pure state of two qubit systems can be written
 as a linear combination in the basis
 $\{\left|\uparrow\uparrow\right\rangle,
    \left|\uparrow\downarrow\right\rangle,
    \left|\downarrow\uparrow\right\rangle,
    \left|\downarrow\downarrow\right\rangle \}$:
 \begin{equation}
 \left|\psi\right\rangle
 = a \left|\uparrow\uparrow\right\rangle
  +b \left|\uparrow\downarrow\right\rangle
  +c \left|\downarrow\uparrow\right\rangle
  +d \left|\downarrow\downarrow\right\rangle.
 \end{equation}
 For two qubit systems,
 a given Hamiltonian $H$ generates the time evolution of the state
 through the Schr\"{o}dinger equation
 $i\hbar\partial_t\left|\psi(t)\right\rangle= H \left|\psi(t)\right\rangle$.
 If the Hamiltonian $H$ is independent of time,
 the time-dependent state is given by
 $\left|\psi(t)\right\rangle
 =\exp\left[-\frac{i H t}{\hbar}\right] \left|\psi(0)\right\rangle
 $,
 where $\left|\psi(0)\right\rangle=\left|\psi\right\rangle$
 is a given state at the initial time.
 By virtue of the unitary transformation $U$ making the Hamiltonian
 diagonal, at time $t$, the state is given by
 \begin{equation}
 \left|\psi(t)\right\rangle
 = G(t)
 \left|\psi(0)\right\rangle,
 \end{equation}
 where the propagator is $G(t)= U \exp\left[-\frac{i U^\dagger H U t}{\hbar}\right]
 U^\dagger$ for the time evolution of the state.

 To quantify entanglement
 for the time evolution of the state,
 we introduce the concurrence as the overlap between the state
 and the spin flipped state at a given time $t$:
 \begin{equation}
  C(|\psi(t)\rangle)
  = \left| \langle\psi(t)\mid \tilde\psi(t)\rangle\right|,
 \end{equation}
 where the spin flipped state is given by
 $|\tilde\psi(t)\rangle
          =\sigma_y \otimes \sigma_y |\psi^*(t)\rangle$ \cite{Wootters}
          with the pauli matrix $\sigma_y$.
 The concurrence ranges from zero (unentangled state)
 to one ( a maximally entangled state).
 To help understanding the entanglement dynamics,
 one can define the overlap between the states
 at the initial time (input state) and at a
 given time $t$ (output state) as the fidelity:
 \begin{equation}
  F(t)
  = \big| \langle\psi(t)\mid \psi(0)\rangle\big|.
 \end{equation}
 If $F(T)=1$, the output quantum state is the same with
 the initial input state at $t=T$, i.e.,
 unentangled (entangled) initial state returns
 to unentangled (entangled) state.
 Then, for time evolution of quantum states,
 entanglement dynamics can be understood from the concurrence and
 fidelity.

 {\it Model}.$-$
 We consider a two superconduting flux quit system where
 the two flux qubits interact each other by controlling
 the Josephson junction energy $E'_J$
 in the superconducting connecting wire in Fig. \ref{fig:1}.
 We assume that the inductances associated with the geometry
 of the system
 is so small that the inductive energy is negligible.
 The Hamiltonian describing the model
 is given by the sum of the charging and Josephson energies:
 \begin{equation}
  H(\{\dot \varphi_i,\dot \varphi',\varphi_i,\varphi'\})
  = H_C(\{\dot \varphi_i,\dot \varphi'\}) + H_J (\{ \varphi_i,\varphi'\}),
 \end{equation}
 where the phases across the Josephson junctions are  $\varphi_i$
 and their time derivatives are  $\dot \varphi_i$.
 The charging energy of Josephson junctions in
 the two qubit loops and the connecting wire is given by
 \begin{equation}
  H_C= \frac{1}{2} \left(\frac{\Phi_0}{2\pi}\right)^2
     \left(
          \sum^2_{i=1} \sum_{\alpha\in\{a,b,c\}}
           C^\alpha_{i}\dot\varphi^{\alpha \ 2}_{i}
   +       C' \dot\varphi'^2 \right),
 \label{CE}
 \end{equation}
 where $C^\alpha(C')$ are the capacitance of the Josephson junctions
 in the qubit (connecting) loops.
 $\Phi_0 = h/2e$ is the unit flux quantum.
 The Josephson energy of the junctions is given by
 \begin{equation}
  H_J = \sum^{2}_{i=1} \sum_{\alpha\in\{a,b,c\}}
          2 E^\alpha_{Ji} \sin^2\frac{\varphi^\alpha_i}{2}
    +     2 E'_J \sin^2\frac{\varphi'}{2},
 \label{JE}
 \end{equation}
 where $E_J$'s are the Josephson energy of junctions
 in the qubit and connecting loops.
 For flux qubits, the charging energy is much smaller
 than the Josephson energy.
 The number of Cooper pairs $n$ and the phase $\varphi$ are
 non-commuting variables, i.e., $[\varphi, n] = i$,
 such that the canonical momentum $P_\varphi$ can be introduced as
 $P_\varphi = n \hbar= -i\hbar\partial_\varphi$,
 where $n = q/2e$ with the charge from the Josephson relation
 $q =C(\Phi_0/2\pi)\dot \varphi$.
 At low energies, then, the charging energy $H_C(\{\varphi_i\})$
 play a role of kinetic energy for a phase particle,
 while the Josephson energy $H_J(\{\varphi_i\})$ plays
 a role of confinement potential for the particle.

 From the fluxoid quantization along the closed paths
 in the two qubit loops and the connecting loop path including
 the two superconducting wires,
 the constraint conditions for the phases are given by
 \begin{eqnarray}
  2\pi ( n_i + f_i ) -\sum_{\alpha \in \{a,b,c\}} \varphi^\alpha_i
   &=& 0, \\
   2\pi n' + (\varphi^a_1-\varphi^a_2) - \varphi' &=& 0,
  \label{constraint}
 \end{eqnarray}
 where $n_i$ and $n'$ are an integer
 and $f_i = \Phi/\Phi_0$ are the applied flux in the qubit loop $i$.
 The constraint of
 $\varphi^a_1$ and $\varphi^a_2 $ in Eq. (\ref{constraint}) induce
 the coupling between the two flux qubits,
 which is call the phase coupling.

 In the low energy limit, generally
 the Hamiltonian of superconducting flux qubit systems
 can be written in terms of the circulating current states in each
 qubit loop \cite{Cho07}.
 For two flux qubit systems, following Ref. \cite{KimCho,Cho07},
 one can write the two-qubit matrix Hamiltonian in terms of
 qubit energy levels, single-qubit tunnelings,
 and two-qubit tunnelings:
 \begin{equation}
 H = \left(
       \begin{array}{cccc}
       E_{\uparrow\uparrow} & -t_{1} & - t_{1}  & -t_{2}^{a} \\
      -t_{1}  & E_{\uparrow\downarrow} &  -t_{2}^{b} & -t_{1} \\
      -t_{1}  & -t_{2}^{b}  & E_{\downarrow\uparrow} & -t_{1}  \\
      -t_{2}^{a}  & -t_{1} &  -t_{1}  & E_{\downarrow\downarrow}
      \end{array} \right),
      \label{Hamiltonian}
\end{equation}
 where $E$'s are the energies for the two qubit states.
 The two qubit states correspond to the local minima
 $\{\varphi^0_{i;m_1,m_2}\}$
 of the Josephson energy $H_J({\varphi_i})$
 with $E^{b,c}_{Ji}=E_{Ji}$ and $\varphi^{b,c}_{i}=\varphi_i$.
 Here, $m_i=\uparrow$ and $\downarrow$ for the qubit $i$.
 The two-qubit level energies are given by
$
 E_{m_1, m_2}
  = \frac{\hbar}{2} \sum_{i=1}^2 \omega_{i;m_1,m_2}
    + H_J(\{\varphi^0_{i;m_1,m_2}\}),
$
 where
 the characteristic oscillating frequencies
 are
 $\omega^2_{i;m_1,m_2}=
 \frac{1}{M_i} \frac{\partial^2}{ \partial\varphi^2_{i}}
 H_J(\{\varphi_i\}) |_{\{\varphi^0_{i;m_1,m_2}\}}$
 with an effective mass
 $M_i = \left(\frac{\Phi_0}{2\pi}\right)^2 C^{(i)}_{\rm eff}$
 and  effective capacitance $C^{(i)}_{\rm eff}$
 in the harmonic oscillator approximation \cite{Orlando}.
 $t_1$ and $t_2$ are the single- and two-qubit tunnelings
 between the two states of two qubits.
 Single-qubit tunneling describes
 single-qubit flipping for
 the macroscopic quantum tunneling
 between the two states of the two qubit states.
 For example,
  $\left|\uparrow\uparrow \right\rangle \Longleftrightarrow
          \left|\downarrow\uparrow \right\rangle$.
 The two-qubit tunneling amplitudes, (i) $t^a_2$ and (ii) $t^b_2$,
 describe the tunneling processes,
 (i) $\left|\uparrow\uparrow \right\rangle \Longleftrightarrow
      \left|\downarrow\downarrow \right\rangle $
 in the parallel pseudo-spin states
 and
 (ii) $\left|\uparrow\downarrow \right\rangle \Longleftrightarrow
      \left|\downarrow\uparrow \right\rangle $
 in the anti-parallel pseudo-spin states.
 The tunneling amplitudes
 are calculated by the numerical methods such as
 WKB approximation, instanton method,
 and Fourier grid Hamiltonian method \cite{Kim03}.

 In fact, the tunneling amplitudes
 and the low energy qubit energies
 are determined by the systems parameters of
 the superconducting flux qubit system.
 Once the parameters are adjusted, generally,
 an artificial spin Hamiltonian
 is given in a form
 from Eq. (\ref{Hamiltonian}):
 \begin{equation}
 H = \sum_{j\in \{1,2\}} \sum_{\alpha\in\{x,y,z\}}
         B^\alpha_j\ S^\alpha_j \
      + \sum_{\alpha\in\{x,y,z\}}  J_\alpha \ S^\alpha_1\
      S^\alpha_2,
      \label{spinHamiltonian}
 \end{equation}
 where $B^x_j = -t_1$, $B^y_j=0$,
 $B^z_1=\left( E_{\uparrow\uparrow} +E_{\uparrow\downarrow}
   -E_{\downarrow\uparrow} -E_{\downarrow\downarrow} \right) /4$,
 $B^z_2=\left( E_{\uparrow\uparrow} -E_{\uparrow\downarrow}
   +E_{\downarrow\uparrow} -E_{\downarrow\downarrow} \right) /4$,
 $J_x =-(t^a_2 + t^b_2 )/2$,
 $J_y = (t^a_2 - t^b_2 )/2$,
 and
 $J_z =\left( E_{\uparrow\uparrow} -E_{\uparrow\downarrow}
   -E_{\downarrow\uparrow} +E_{\downarrow\downarrow} \right) /4 $.
 The single qubit tunnelings play the role of a transverse
 magnetic field while the energy difference of two-qubit levels
 correspond to the applied magnetic field parallel to the $z$-direction of
 spins.
 Note that the $x$- and $y$-components of the exchange interaction
 are determined by the two-qubit tunnelings and the $z$-component
 of the interaction is the energy difference between
 the parallel spin state and the anti-parallel spin state.
 Consequently, Eq. \ref{spinHamiltonian}
 shows that an XYZ quantum spin model with magnetic fields
 can be realizable in any two flux qubit system \cite{Cho07}.
 We will discuss a specific realization of a quantum spin model
 with adjusted system parameters and entanglement dynamics
 for an input state.

 {\it Realizable artificial spin systems.}
 Case I.
 For $E_J'=0.0 E_J$ and $E_{J1}=0.7E_J$,
 a two-spin Hamiltonian can be constructed by
 the relations of
 $ E_{\uparrow\uparrow} =E_{\uparrow\downarrow}
   =E_{\downarrow\uparrow} =E_{\downarrow\downarrow}$
 and $t^a_2=t^b_2=t_2$.
 The numerical values of the macroscopic quantum tunnelings
 are obtained as  $t_1=0.0075E_J$ and $t_2^{a(b)}=0.00024E_J$.
 From Eq. \ref{spinHamiltonian},
 the two flux qubit system is described by the corresponding
 spin Hamiltonian:
 \begin{equation}
 H =  J \ S^x_1\ S^x_2 + B \ (S^x_1 + S^x_2 ),
 \end{equation}
 where $B = -t_1=$ and $J =-t_2$.
  Note that the entanglement dynamics of this spin system
 is determined only by the single- and two-qubit tunneling amplitudes.

 For the spin system, the concurrence is given by
 \begin{equation}
 C(t)=\Big[ C_0+C_1\cos{4 J t} \Big]^{1/2},
 \end{equation}
 where $C_0=[(a+d)^2-(b+c)^2]^2/4 + [(a-d)^2-(b-c)^2]^2/4$ and
 $C_1=-[(a+d)^2-(b+c)^2][(a-d)^2-(b-c)^2]/2 $.
 It should be noticed that the concurrence does not depend on
 the magnetic field $B=-t_1$, i.e., the single-qubit tunneling.
 The concurrence is an oscillating function
 with respect of the exchange interaction $J=-t_2$
 with the characteristic period of time $T = \pi/2J$.
 This shows that any entangled state cannot be generated by applying
 the magnetic field in this quantum spin system.
 At $t=(m+1) \pi/2J$ with an integer $m$,
 the concurrence reaches its maximum value $C(t)=2|ad-bc|$, i.e.,
 a maximally entangled state.
 At $t=\frac{1}{4J} \cos^{-1}\frac{C_0}{C_1}$, the entanglement
 disappears, i.e., the input state involves to be unentangled.

 The fidelity of this quantum spin system is given by
 \begin{equation}
 F(t)
  =\left[ F_0 +  \sum_{\sigma=\pm} F^{\sigma}_1\cos 2(B+\sigma J) t
    + F_2 \cos 4 B t
 \right]^{1/2},
 \end{equation}
 where
 $F_0=1-[(a+d)^2+(b+c)^2] [(a-d)^2+(b-c)^2]/2-[(a+d)^2-(b+c)^2]^2/8$,
 $F^{+}_1=(a+b+c+d)^2[(a-d)^2+(b-c)^2]/4$,
 $F^{-}_1=(a-b-c+d)^2[(a-d)^2+(b-c)^2]/4$,
 and $F_2=[(a+d)^2-(b+c)^2]^2/8$.
 For zero magnetic field, the quantum state
 evolves in time due to the exchange interaction.
 The fidelity has twice longer period of time than the concurrence.

 Let us study entanglement when the initial input states
 are not in an entangled state.
 We choose the case of $c=d=0$,
 i.e., $\left|\Psi(0)\right\rangle
 =a\left|\uparrow\uparrow\right\rangle
 +b\left|\uparrow\downarrow\right\rangle$, in which
 the initial state is a product state.
 It is clearly shown that the exchange interaction
 makes the artificial spins entangled in the concurrence
 $C(t)=|(2a^2-1)\sin 2 J t|$.
 The period of the concurrence is $T=\pi/J$.

 For $a=d$ and $b=c$, the initial state can be written
 in the Bell basis $\{ \Psi^\pm, \Phi^\pm\}$
 with $\Psi^\pm=\frac{1}{\sqrt{2}}
 ( \left|\uparrow\downarrow\right\rangle \pm
   \left|\downarrow\uparrow\right\rangle )$
   and
   $\Phi^\pm=\frac{1}{\sqrt{2}}
 ( \left|\uparrow\uparrow\right\rangle \pm
   \left|\downarrow\downarrow\right\rangle )$.
 The state is given by
  $\left|\psi(t)\right\rangle
 = \sqrt{2} ( a \left|\Phi^+\right\rangle
 + b \left|\Psi^+\right\rangle)$.
 Initially the input state is an entangled
 state quantified by the concurrence $C(0)=2 |a^2-b^2|$.
 For the time evolution of the state,
 the concurrence is not a function of time,
 i.e., $C(t)=2|a^2-b^2|$.
 However,
 the fidelity is a function of time.
 It does not depend on
 the exchange interaction, i.e.,
 $F(t) =\sqrt{A_0 + A_1 \cos 4 B t }$,
 where $A_0=1-2 (a^2-b^2)^2$ and $A_1=2 (a^2-b^2)^2$.
 This can be understood as follows.
 The initial state can be rewritten
 in the eigen basis:
   $\left|\psi(t)\right\rangle
   = (a+b)\left|\psi_0\right\rangle
     +(a-b)\left|\psi_3\right\rangle$,
     where $\left|\psi_0\right\rangle$ and
     $\left|\psi_3\right\rangle$ are the ground
     state and the third excited state, respectively.
 At a given time $t$, then,
   $\left|\psi(t)\right\rangle
   = (a+b)e^{-iE_0t} \left|\psi_0\right\rangle
     +(a-b)e^{-iE_3 t}\left|\psi_3\right\rangle$,
 where the energies for the ground state and he third excited
 state are $E_0=J-2 B$ and $E_3=J+2 B$.
 As a result, for the constant concurrence, the fidelity
 is oscillating in time.

 \begin{figure}
 \begin{overpic}
 [width=1.6in]{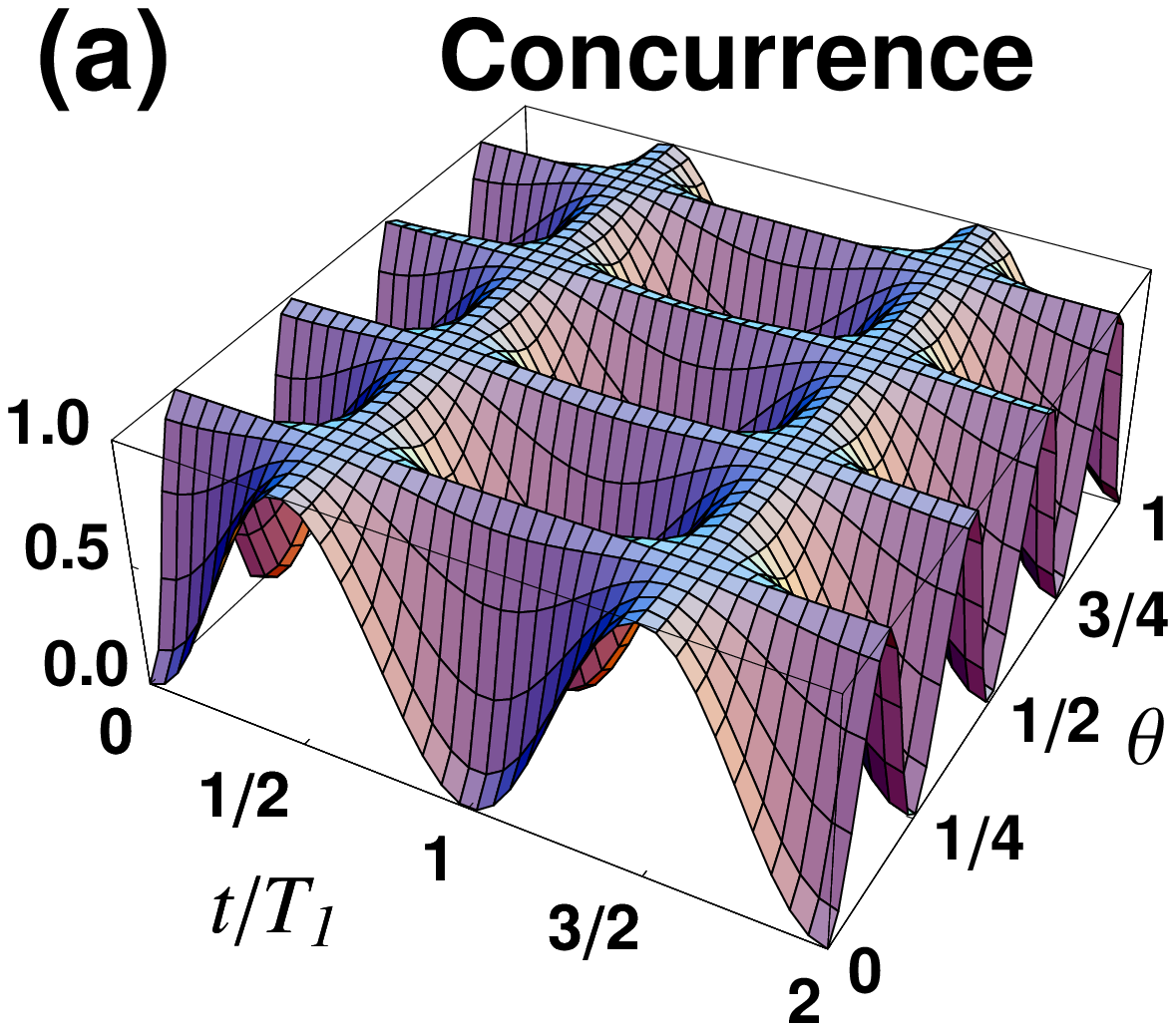}
 \end{overpic}
 \includegraphics[width=1.6in]{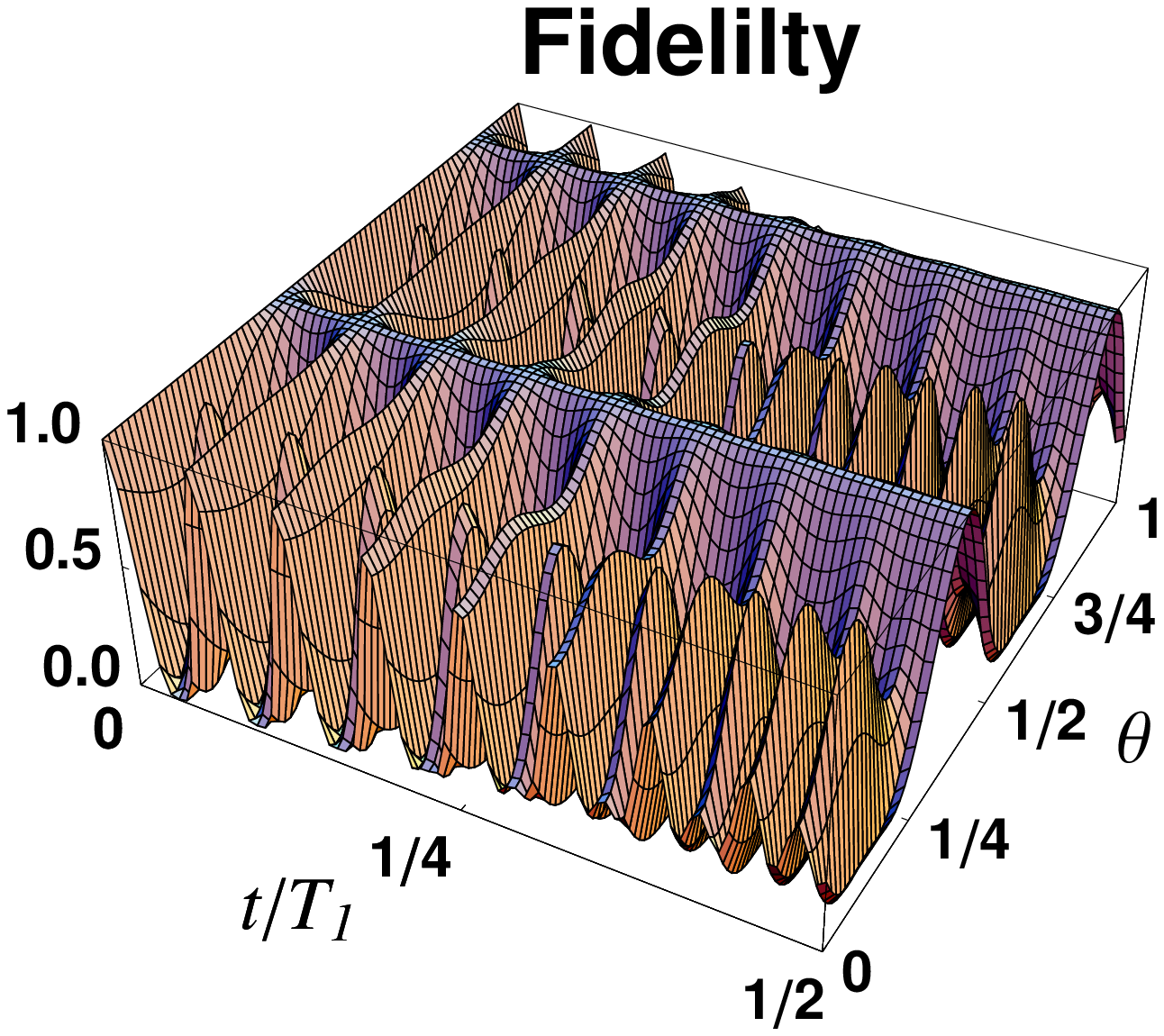}
   \begin{overpic}
  [width=1.6in]{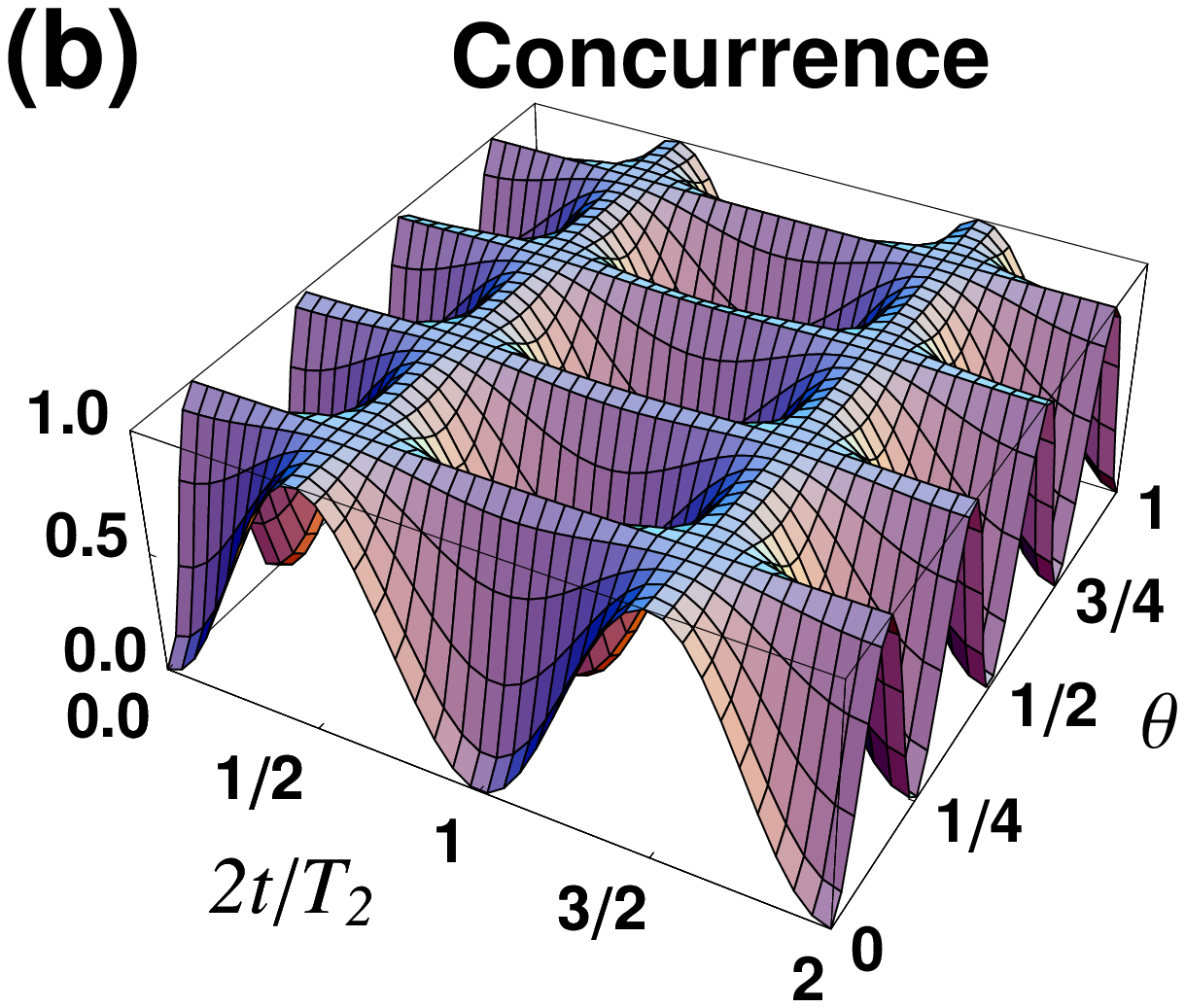}
  \end{overpic}
  \includegraphics[width=1.6in]{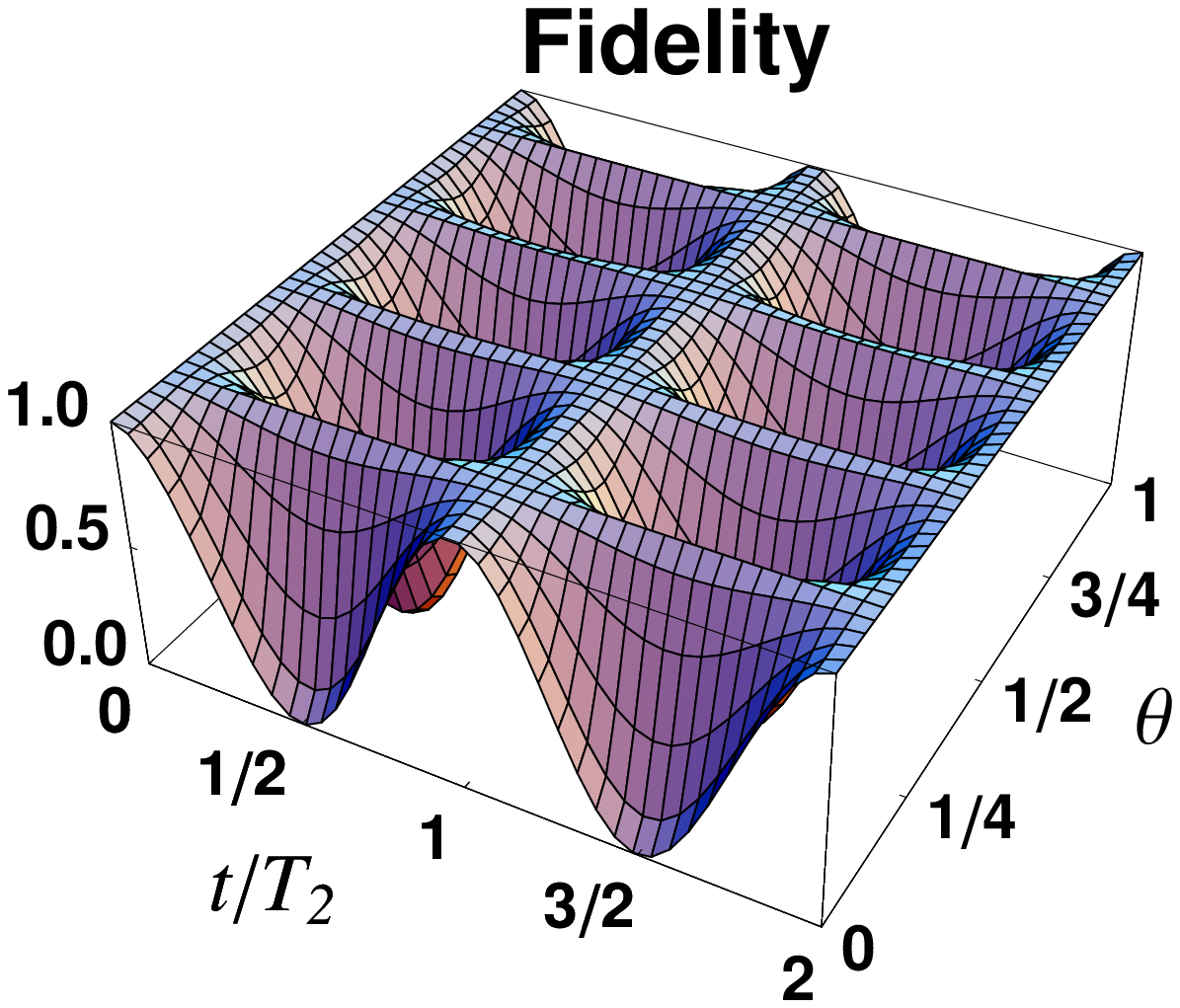}
    \begin{overpic}
  [width=1.6in]{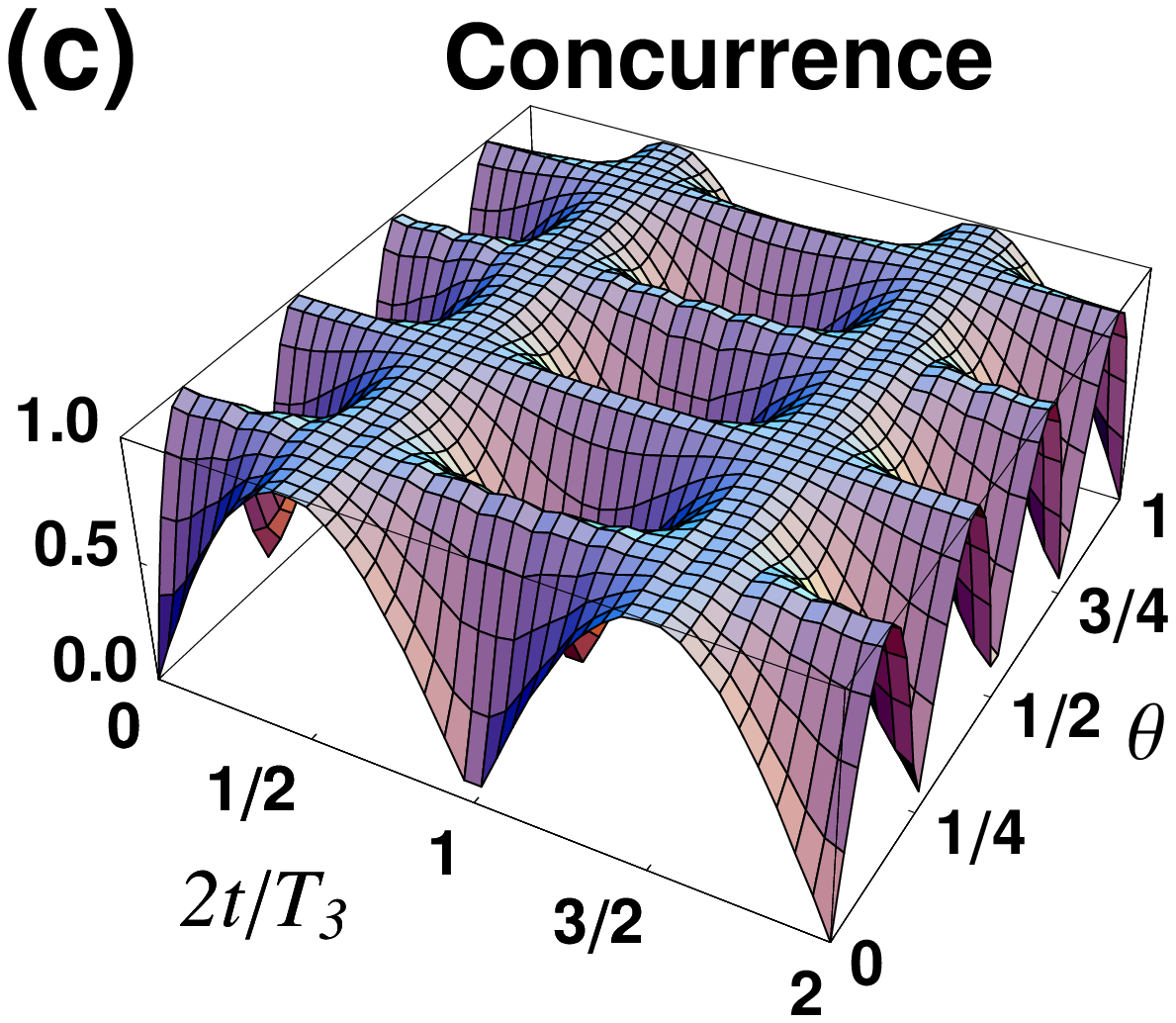}
  \end{overpic}
  \includegraphics[width=1.6in]{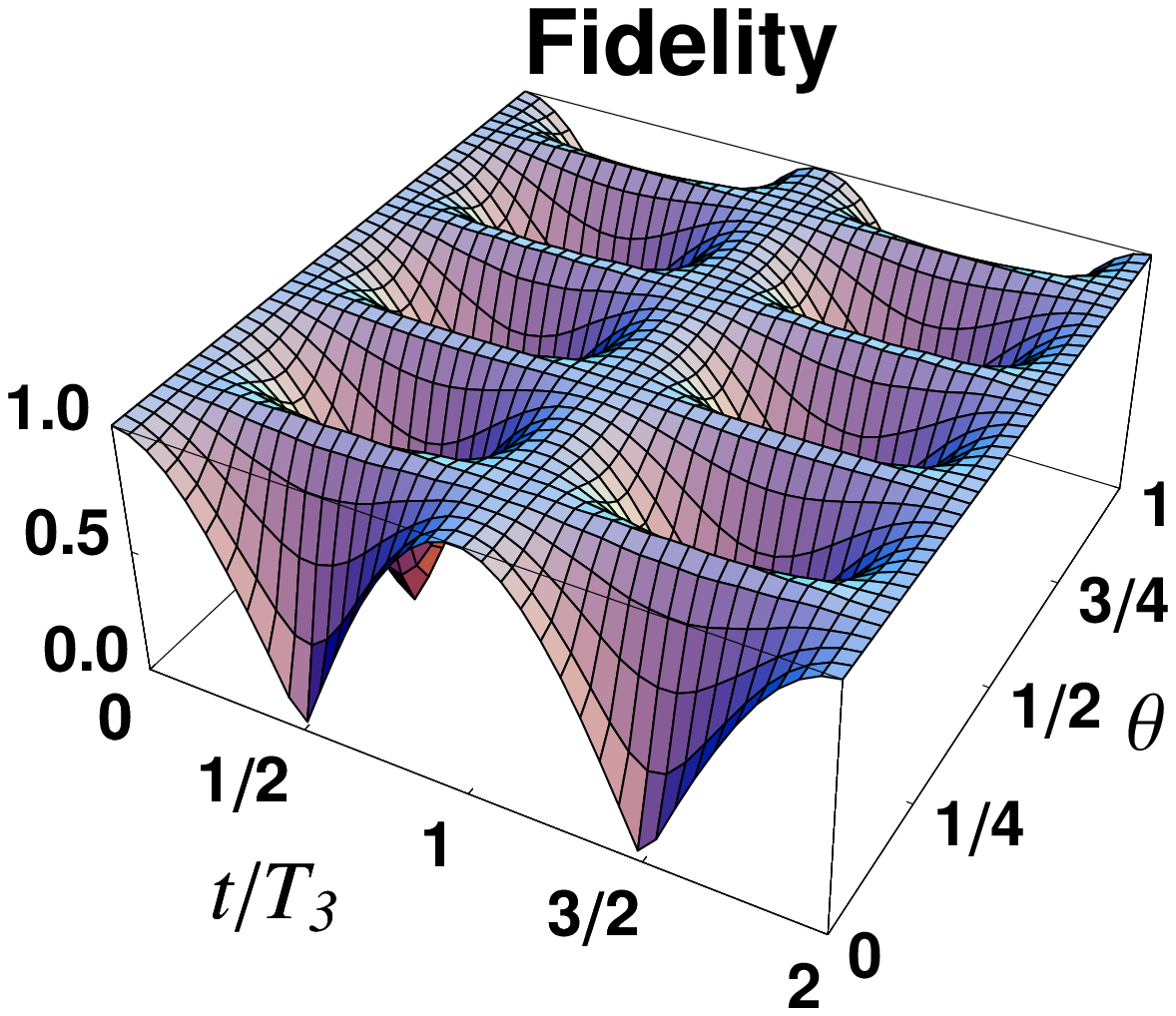}
 \caption{(color online)
  Time evolutions of the concurrence $C(t,\theta)$ and
  the fidelity  $F(t,\theta)$
  for the input state $\left|\Psi(0)\right\rangle
     =  \cos 2\pi\theta \left|\uparrow\uparrow\right\rangle
      + \sin 2\pi\theta \left|\downarrow\downarrow\right\rangle$
  at the co-resonance point $f_1=f_2=0.5$
  in the two superconducting flux quit system.
 (a) Case I.
 For $E'_J=0.0E_J$ and $E_{J1}=0.7E_J$,
 the two flux qubit system corresponds to the spin Hamiltonian
 $  H =  J \ S^x_1\ S^x_2 + B \ (S^x_1 + S^x_2 )$
 with the exchange interaction  $J=-t^a_2$ and the magnetic field $B=-t_1$.
 The single- and two-qubit tunneling amplitudes
 are given by $t_1=0.0075E_J$ and $t_2=0.00024E_J$.
 The characteristic period of time  is $T_1=\pi/2J$
 for the both concurrence and fidelity.
 (b) Case II.
 For $E'_J=0.6E_J$ and $E_{J1}=0.7E_J$,
 the two flux qubit system maps into the spin Hamiltonian
 $H = J\ ( S^x_1\ S^x_2 - \ S^y_1\ S^y_2 )
    + J_z \ S^z_1\ S^z_2$
 with the exchange interaction $J=-t^a_2/2$
 and $J_z=(E_{\uparrow\uparrow}-E_{\uparrow\downarrow})/2$.
 The two-quit tunneling amplitude is $t^a_2=0.00024E_J$ and
 the energy difference between the two states is $J_z=-0.425045E_J$.
 The period of time is $T_2=\pi/2J$.
 The fidelity has twice period of the concurrence.
 (c) Case III.
 For $E'_J=0.05E_J$ and $E_{J1}=0.7E_J$,
 The two flux qubit system is described by the spin Hamiltonian
 $ H =  J ( \ S^x_1\ S^x_2 - \ S^y_1\ S^y_2 )
        + J_z \ S^z_1\ S^z_2+ B \ (S^x_1 + S^x_2 )$
 with the magnetic field $B=-t_1$
 and the exchange interactions  $J=-t^a_2/2$,
 and $J_z=(E_{\uparrow\uparrow}-E_{\uparrow\downarrow})/2$.
 The single- and two-qubit tunneling amplitudes are
 $t_1=0.0024E_J$ and $t^a_2=0.00024E_J$ and
 the energy difference becomes $J_z=-0.05E_J$.
 The characteristic period of the time evolutions
 is $T_3=0.68 \pi/2J$.
 The fidelity has twice period of the concurrence.
 }
 \label{fig:2}
 \end{figure}

 Case II.
 For $E'_J=0.6 E_J$ and $E_{J1}=0.7E_J$,
 one find the relations:
   $ E_{\uparrow\uparrow}=E_{\downarrow\downarrow}$
  and
   $E_{\downarrow\uparrow}=E_{\uparrow\downarrow}$,
   and $t_1= t^b_2=0$.
 The corresponding two spin system
  can be written as
 \begin{equation}
 H = J\ ( S^x_1\ S^x_2 - \ S^y_1\ S^y_2 )
    + J_z \ S^z_1\ S^z_2 ,
 \end{equation}
 where  $J =-t^a_2/2$,
 and $J_z = (E_{\uparrow\uparrow} -E_{\uparrow\downarrow})/2$.
 The numerical values of the two-quit tunneling amplitude
 and the energy difference between the two states
 are given by  $t^a_2=0.00024E_J$ and $J_z=-0.425045E_J$.
 The Hamiltonian describe an anisotropic spin exchange interaction
 belonging in the class of the XYZ spin model.
 Interestingly, the $x$- and
 $z$-components of the interaction are anti-ferromagnetic
 because $J < 0$ and $J_z <0$
 while the $y$-component is ferromagnetic. To our knowledge,
 it is unlikely to find a class of spin Hamiltonian naturally.

 For the spin Hamiltonian, the concurrence is given by
 \begin{equation}
 C(t)
  =\left[ C_0 + \sum_{\sigma=\pm} C^\sigma_1\cos 4(J+\sigma J_z) t
   + C_2 \cos 8 J t
 \right]^{1/2},
 \end{equation}
 where $C_0=[(a+d)^4+(a-d)^4]/4+4b^2c^2$,
 $C^+_1=-2(a+d)^2bc$, $C^-_1=2(a-d)^2 bc$, and $C_2=-(a^2-d^2)^2/2$.
 The fidelity is given by
 \begin{equation}
 F(t)
  =\left[ F_0 +  \sum_{\sigma=\pm}F_1\cos 2(J+\sigma J_z) t
   + F_3 \cos 4 J t
   \right]^{1/2},
 \end{equation}
 where
 $F_0=1-2(a^2+d^2)(b^2+c^2)-(a^2-d^2)^2/2$, $F^+_1=(a+d)^2 (b^2+c^2)$,
 $F^-_1=(a-d)^2(b^2+c^2)$,
 and $F_2=(a^2-d^2)^2/2$.
 This shows that
 the concurrence and fidelity have a similar dynamic property.
 However, the fidelity has twice longer period
 than the concurrence.

 Compared to the Case I,
 for the initial product state $\left|\Psi(0)\right\rangle
 =a\left|\uparrow\uparrow\right\rangle
 +b\left|\uparrow\downarrow\right\rangle$,
 the concurrence becomes
 $C(t)=a^2|\sin 4 J t|$ with the period $T=\pi/2J$.
 The input state
 $\left|\psi(t)\right\rangle
 = \sqrt{2} ( a \left|\Phi^+\right\rangle
 + b \left|\Psi^+\right\rangle)$
 evolutes
 and
 its concurrence
 is oscillating with
 $C(t)=2\sqrt{a^4+b^4-2a^2b^2\cos 4(J-J_z) t}$.
 Also, the fidelity is given by
 $F(t) =\sqrt{A_0 + A_1 \cos 4 J t }$,
 where $A_0=1-8 a^2 b^2$ and $A_1=8 a^2 b^2$.
 Thus, the Case I and II show a different characteristic
 entanglement dynamics depending on
 the realizable two artificial spin models
 in the flux qubit systems.

 Case III.
 For $E'_J=0.05E_J$ and $E_{J1}=0.7E_J$,
 The two-qubit energies have
 the relations:  $ E_{\uparrow\uparrow}=E_{\downarrow\downarrow}$,
 and  $E_{\downarrow\uparrow}=E_{\uparrow\downarrow}$.
 The two-qubit tunneling becomes $t^b_2=0$.
 We find another realization of a two spin system:
 \begin{equation}
 H =  J \ (S^x_1\ S^x_2 - \ S^y_1\ S^y_2 )
        + J_z \ S^z_1\ S^z_2+ B \ (S^x_1 + S^x_2 ),
 \end{equation}
 where $B = -t_1$, $J =-t^a_2/2$,
 and $J_z = (E_{\uparrow\uparrow} -E_{\uparrow\downarrow})/2$.
 For the system parameters of the superconducting flux qubits,
 the single- and two-qubit tunneling amplitudes are
 $t_1=0.0024E_J$ and $t^a_2=0.00024E_J$ and
 the energy difference is $J_z=-0.05E_J$.
 The expressions of the concurrence and fidelity are too lengthy to display.

 In Fig. \ref{fig:2},
 we plot the concurrences and fidelities
 as a function of time $t$ and the initial state parameter $\theta$
 to give the comparison of entanglement dynamics between
 the three different spin models
 for the same initial state
 $\left|\Psi(0)\right\rangle
     =  \cos 2\pi\theta \left|\uparrow\uparrow\right\rangle
      + \sin 2\pi\theta \left|\downarrow\downarrow\right\rangle$.
 Explicitly, the different values of system parameters
 controlling the two flux qubits
 are given in the captions of the figures.
 For the time evolution of the initial state,
 it is shown that
 unentangled (entangled) state can become entangled (unentangled)
 state even though the specification of the superconducting devices
 are different each other.

 {\it Summary.}
 A two superconducting flux qubit system has been considered to
 investigate a possible realization of quantum spin models.
 Three different artificial spin models were demonstrated
 by varying controllable system parameters.
 The realizable spin models in the flux qubit system
 are not likely to find naturally.
 We discussed the entanglement dynamics of the artificial spin models
 in the specific parameter values of the two superconducting
 flux qubit system.
 It was found that the input
 unentangled (entangled) state can become
 maximally entangled (unentangled) state
 irrespective of
 the specifications of the superconducting devices.
 Such a maximally entangled state should be observable
 experimentally.

  We thank Huan-Qiang Zhou and
  John Paul Barjaktarevic for helpful
  discussions.

\end{document}